\documentclass[journal=jacsat,manuscript=article]{achemso}

\usepackage{chemformula} 
\usepackage[T1]{fontenc} 
\newcommand{\upcite}[1]{\textsuperscript{\textsuperscript{\cite{#1}}}}

\author{Zhaoqing Ding}
\affiliation{Beijing National Laboratory for Condensed Matter Physics and Institute of Physics, Chinese Academy of Sciences, Beijing 100190, China}
\alsoaffiliation[Second University]
{School of Physical Sciences, University of Chinese Academy of Sciences, Beijing 100049, China}

\author{Yongjie Xie}
\affiliation{Beijing National Laboratory for Condensed Matter Physics and Institute of Physics, Chinese Academy of Sciences, Beijing 100190, China}
\alsoaffiliation[Second University]
{School of Physical Sciences, University of Chinese Academy of Sciences, Beijing 100049, China}

\author{Xuejiao Chen}
\affiliation{School of Photoelectric Engineering, Changzhou Institute of Technology, Changzhou, Jiangsu 213002, China}

\author{Sheng Wang}
\affiliation{Beijing National Laboratory for Condensed Matter Physics and Institute of Physics, Chinese Academy of Sciences, Beijing 100190, China}
\alsoaffiliation[Second University]
{School of Physical Sciences, University of Chinese Academy of Sciences, Beijing 100049, China}

\author{Zhen Wang}
\affiliation{Beijing National Laboratory for Condensed Matter Physics and Institute of Physics, Chinese Academy of Sciences, Beijing 100190, China}
\alsoaffiliation[Second University]
{School of Physical Sciences, University of Chinese Academy of Sciences, Beijing 100049, China}
\alsoaffiliation[Third University]
{Institute of High Energy Physics, Chinese Academy of Sciences, Beijing 100049, China}

\author{Zeguo Lin}
\affiliation{Beijing National Laboratory for Condensed Matter Physics and Institute of Physics, Chinese Academy of Sciences, Beijing 100190, China}
\alsoaffiliation{School of Physical Sciences, University of Chinese Academy of Sciences, Beijing 100049, China}

\author{Enling Wang}
\affiliation{Beijing National Laboratory for Condensed Matter Physics and Institute of Physics, Chinese Academy of Sciences, Beijing 100190, China}
\alsoaffiliation{School of Physical Sciences, University of Chinese Academy of Sciences, Beijing 100049, China}

\author{Xiaofeng Wu}
\affiliation{Beijing National Laboratory for Condensed Matter Physics and Institute of Physics, Chinese Academy of Sciences, Beijing 100190, China}
\alsoaffiliation{School of Physical Sciences, University of Chinese Academy of Sciences, Beijing 100049, China}

\author{Mingyu Yang}
\affiliation{Beijing National Laboratory for Condensed Matter Physics and Institute of Physics, Chinese Academy of Sciences, Beijing 100190, China}
\alsoaffiliation{School of Physical Sciences, University of Chinese Academy of Sciences, Beijing 100049, China}

\author{Yuelong Xiong}
\affiliation{Beijing National Laboratory for Condensed Matter Physics and Institute of Physics, Chinese Academy of Sciences, Beijing 100190, China}
\alsoaffiliation{School of Physical Sciences, University of Chinese Academy of Sciences, Beijing 100049, China}

\author{Meng Meng}
\affiliation{Beijing National Laboratory for Condensed Matter Physics and Institute of Physics, Chinese Academy of Sciences, Beijing 100190, China}

\author{Fang Yang}
\affiliation{Beijing National Laboratory for Condensed Matter Physics and Institute of Physics, Chinese Academy of Sciences, Beijing 100190, China}

\author{Jiandi Zhang}
\affiliation{Beijing National Laboratory for Condensed Matter Physics and Institute of Physics, Chinese Academy of Sciences, Beijing 100190, China}
\alsoaffiliation{School of Physical Sciences, University of Chinese Academy of Sciences, Beijing 100049, China}

\author{Xianggang Qiu}
\affiliation{Beijing National Laboratory for Condensed Matter Physics and Institute of Physics, Chinese Academy of Sciences, Beijing 100190, China}
\alsoaffiliation{School of Physical Sciences, University of Chinese Academy of Sciences, Beijing 100049, China}
\email{xgqiu@iphy.ac.cn}

\author{Xiaoran Liu}
\affiliation{Beijing National Laboratory for Condensed Matter Physics and Institute of Physics, Chinese Academy of Sciences, Beijing 100190, China}
\email{xiaoran.liu@iphy.ac.cn}

\author{Jiandong Guo}
\affiliation{Beijing National Laboratory for Condensed Matter Physics and Institute of Physics, Chinese Academy of Sciences, Beijing 100190, China}
\alsoaffiliation{School of Physical Sciences, University of Chinese Academy of Sciences, Beijing 100049, China}
\email{jdguo@iphy.ac.cn}

\title[An \textsf{achemso} demo]
  {Emergent Chiral Spin Crystal Phase in (111) SrRuO$_3$ Thin Films}

\abbreviations{IR,NMR,UV}

\begin{document}







\begin{abstract}
Perovskite ruthenates are fascinating playgrounds for exploring topological spin textures, but generally rely on extrinsic mechanisms to trigger the noncoplanar states. Here we report the discovery of an emergent chiral spin crystal phase in (111) SrRuO$_3$ epitaxial films, characterized by a significant topological Hall effect and noncoplanar spin arrangements with different propagation vectors along two orthogonal directions. Instead of driven by the enhanced Dzyaloshinskii-Moriya interaction due to broken inversion symmetry at heterointerfaces, this emergent state arises intrinsically from the interplay of dipolar interactions and magnetic frustration, leading to the stabilization of topological phases in much thicker films. These findings open a new pathway for creating and controlling the topological spin states in perovskites, with broad implications for spintronic device design.
\end{abstract}

\section{Introduction}
The emergence of noncoplanar spin textures in magnetic materials often activates the chirality degrees of freedom via the adjacent three-spin scalar product, $\chi = \vec{S_i} \cdot (\vec{S_j} \times \vec{S_k})$, giving rise to fertile topologically nontrivial states.\upcite{Tokura_2021_CR,Kurumaji_2019_Science,Simonet_2012_EPJST,Wen_1989_PRB,Okubo_2012_PRL} These include the magnetic quasi-particles (e.g., skyrmions and merons) with definitive topological winding numbers, and periodic noncoplanar spin textures with multiple propagation $Q$ vectors along different directions --- magnetic spin crystals.\upcite{Rossler_2006_Nature,Leonov_2015_NC,Heinze_2011_NP,Ezawa_2010_PRL,Rohart_2013_PRB,Gobel_2021_PR,Hayami_2021_JPCM} The resultant nonzero scalar chiralities generate an additional contribution to the Hall effect through the spin Berry phase, termed as the topological Hall effect (THE), which is generally regarded as the hallmark of chiral spin states \upcite{Taguchi_2001_Science,Onoda_2003_PRL,Kanazawa_2011_PRL,Kalitsov_2009_JPCS,Denisov_2018_PRB,Bruno_2004_PRL,Chen_2023_AFM}. These fascinating quantum states of matter have been innovating the understanding on the topological and geometrical aspects of correlated systems, and exhibiting potential applications in the fields of spintronics, next-generation memory devices, and quantum computing.\upcite{Zhou_2024_AM,Zhang_2020_JPCM,Zhang_2020_NSR}

Recently, the perovskite ruthenate SrRuO$_3$ has attracted significant attention for investigating topological spin states.\upcite{Seddon_2021_NC,Matsuno_2016_SA,Lindfors_2017_PSSB,Wang_2018_NM,Wang_2020_AM,Wang_2019_NM}  Bulk SrRuO$_3$ is a 4{\it d} itinerant ferromagnet below its Curie temperature ($T_C$) around 150 K, with a saturated magnetization of $\sim$1.6 $\mu_B$ per Ru$^{4+}$ ion.\upcite{Koster_2012_RMP,Klein_1996_PRL} Based on numerous experimental efforts, a seemingly generic design paradigm has been established. Specifically, heterostructures composed of (001)-oriented ultrathin SrRuO$_3$ slab of only several unit cells (u.c.) thickness and oxides with strong spin-orbit coupling or ferroelectric oxide layers were constructed to induce substantial Dzyaloshinskii-Moriya (DM) interaction at the interface, which is key to triggering noncoplanar spin textures in SrRuO$_3$. Following this strategy, magnetic skyrmion phases have been reported in SrRuO$_3$/SrIrO$_3$,\upcite{Matsuno_2016_SA} SrRuO$_3$/BaTiO$_3$,\upcite{Wang_2018_NM} SrRuO$_3$/BiFeO$_3$ \upcite{Wang_2020_AM} systems, and an incommensurate spin crystal phase in SrRuO$_3$/PbTiO$_3$ system.\upcite{Seddon_2021_NC}

However, several limitations and issues are brought about in this paradigm. First, the DM interaction as the driving force of noncoplanar spin textures is extrinsically induced from interface, whose decaying magnitude leads to the formation of compelling magnetic phases that are fragile and stabilized only within very few unit cells of SrRuO$_3$. In addition, for SrRuO$_3$ thin films below a thickness of $\sim$10 u.c. (about 4 nm), the $T_C$ drops rapidly until the complete suppression of ferromagnetism near the two-dimensional limit. As the result, the associated topologically nontrivial spin states also emerge at rather low temperatures.\upcite{Seddon_2021_NC,Matsuno_2016_SA,Lindfors_2017_PSSB,Wang_2018_NM,Wang_2020_AM,Wang_2019_NM} On the other hand, it also calls into question the topological interpretation of the THE observed in ultrathin SrRuO$_3$ slabs, because inhomogeneous issues caused by defects,\upcite{Skoropata_2021_PRB,Tian_2022_APL} thickness variation,\upcite{WangL_2020_NL,Kimbell_2020_RRM,Wysocki_2020_PRM} or structural modification at the interfaces \upcite{Qin_2019_AM} are inevitably magnified, such that the topological Hall signal may be alternatively interpreted as a summation of multiple anomalous Hall signals resulting from different types of inhomogeneity.\upcite{Kimbell_2022_CM} All these highlight the need for more intrinsic paradigm to developing robust topological spin states in SrRuO$_3$. 

\begin{figure}[htp]
\centering
\includegraphics[width=0.95\textwidth]{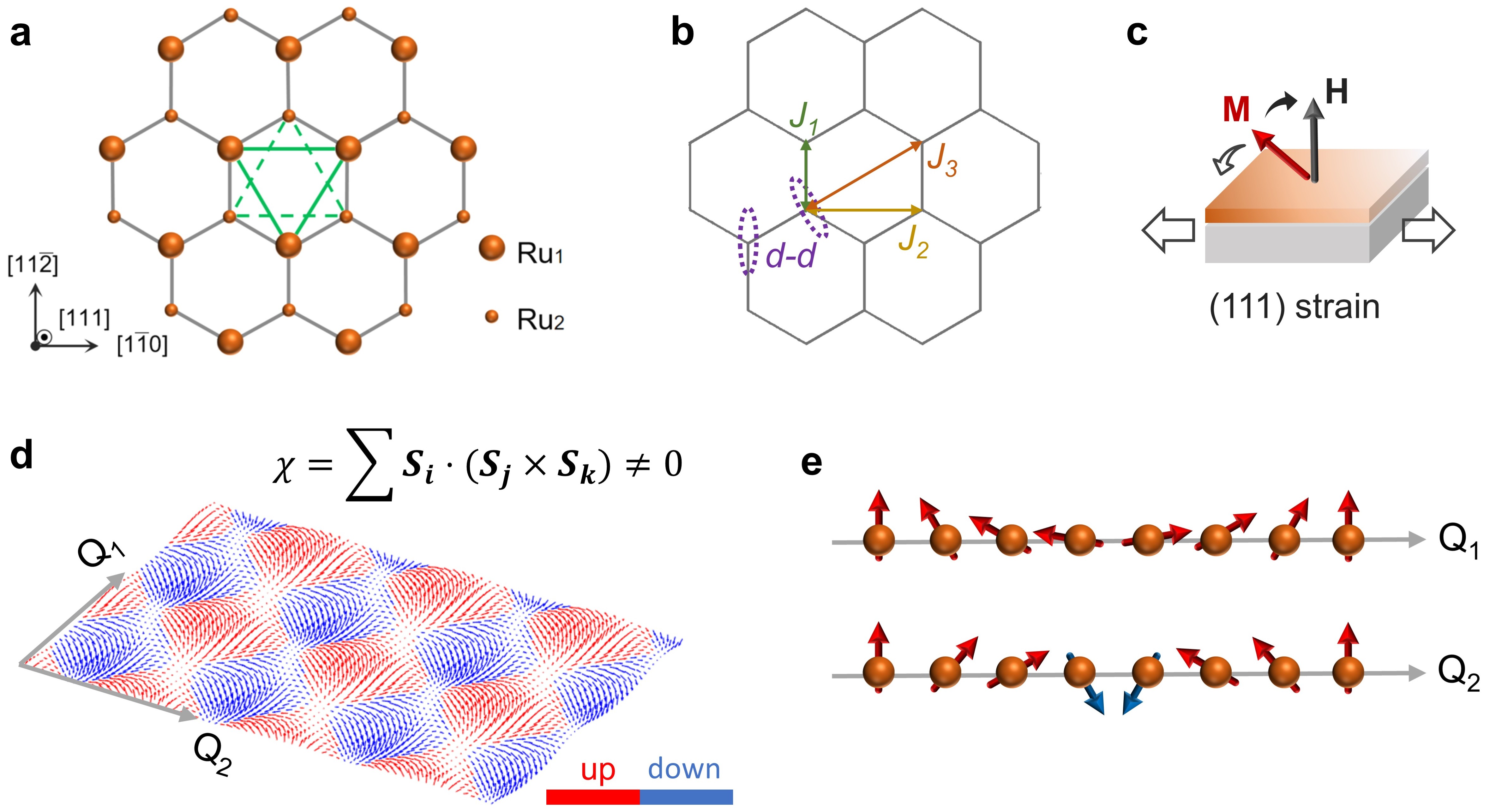}
\caption{\label{Fig1} 
{\bf Schematic of the chiral spin crystal phase. (a) The buckled honeycomb lattice constructed by two neighboring Ru layers of SrRuO$_3$ along the [111] direction. Ru$_1$ and Ru$_2$ refer to the Ru atoms of respective layers. (b) The dipolar interaction and the nearest-neighbor interactions up to the third order. (c) Illustration of the competition between Zeeman energy and tensile-strain induced magnetic anisotropy. (d) Overview of the double-$Q$ chiral spin crystal phase on the buckled honeycomb lattice. (e) Schematics of the cycloidal spin arrangements along two orthogonal $Q_1$ and $Q_2$ directions.}}
\end{figure}

While the honeycomb-related lattices being the heart of exploring topological electronic states,\upcite{Haldane_1988_PRL,Xiao_2011_NC,Ruegg_2013_PRB,Si_2017_PRL,Marthinsen_2018_PRM,Wang_2022_CPB,Lin_2021_AM} it has been lately recognized as a promising playground for topological spin states as well, since the coupled spin and charge degrees of freedom of electrons can lead to strong frustration due to the competition among anisotropic exchange interactions and multiple-spin interactions.\upcite{Hayami_2021_JPCM} Notably, honeycomb-like lattice motifs are naturally established by stacking a perovskite along the [111] direction (e.g., when viewed along the [111] direction of SrRuO$_3$, two neighboring Ru atomic planes set in a buckled honeycomb lattice [Figure~\ref{Fig1}a-b]). 

In this Letter, we report on the experimental realization of a chiral spin crystal phase in (111)-oriented SrRuO$_3$ epitaxial thin films [Figure~\ref{Fig1}c-e]. Remarkable and robust THE was observed right below the Curie temperature at 152 K, within a magnetic field range of 2.5 - 6 T. Concurrently with the THE, magnetic force microscopy (MFM) imaging provided direct real-space evidences on the formation of periodic noncoplanar spin textures with propagation vectors along two directions, denoted as ``double-$Q$'' chiral spin crystal. The exotic spin structures can be reasonably described as a superposition of two orthogonal cycloidal spin spirals. Theoretical micromagnetic simulations corroborate the driving force is not from DM interaction associated with interfacial broken inversion symmetry but from frustrated interactions given by the (111) lattice geometry, and reveal the essence of dipolar interaction and higher-order anisotropic interactions as decisive parameters to this emergent spin state.\\ 

\section{Results and Discussion}

Epitaxial (111) SrRuO$_3$ thin films ($\sim$20 nm) were fabricated using the pulsed laser deposition technique on TbScO$_3$ substrates. Structural characterizations demonstrate the high crystallinity of films with coherent strain status. Notably, previous studies have revealed a tendency of (111) SrRuO$_3$ epitaxial films towards the easy-in-plane magnetic anisotropy as the magnitude of tensile strain reaches +1.5\% on KTaO$_3$ substrate.\upcite{Rastogi_2019_APLM,Ding_2023_npj} Here, TbScO$_3$ provides a moderate tensile strain of +0.5\%, leading to a magnetic easy-axis $\sim$75$^{\circ}$ tilted from the film normal and projected along the [-101] direction in the (111) plane. The temperature dependence of resistivity $\rho$(T) indicates a metallic behavior for the entire measured temperature range with a `kink' near 152 K, referring to the para- to ferro-magnetic transition [see Supporting Information Figure S1 for more details].   

\begin{figure}[htp]
\centering
\includegraphics[width=0.95\textwidth]{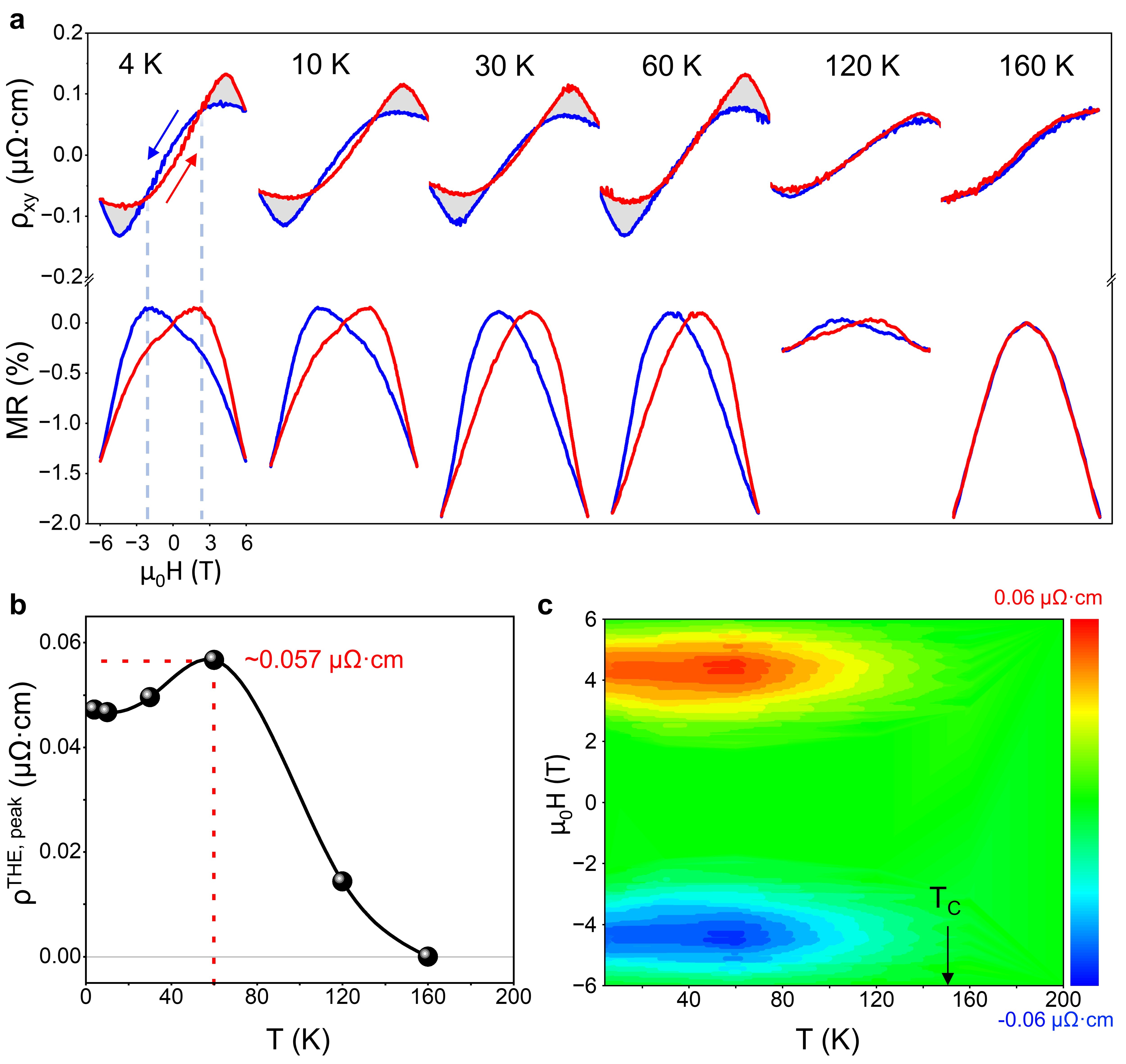}
\caption{\label{Fig2} 
{\bf THE in (111) SrRuO$_3$ thin films. (a) The Hall resistivity $\rho_{xy}$ and the longitudinal magnetoresistance MR at different temperatures across T$_C$. The blue curves refer to scans recorded by sweeping magnetic field from positive to negative, whereas the red ones from negative to positive. The hump-like THE contribution at each temperature is shadowed in gray. (b) The peak value of THE as a function of temperature. (c) Mapping of the extracted topological Hall resistivity as a function of temperature and magnetic field.}}
\end{figure}

We first discuss the THE results. Figure~\ref{Fig2}a exhibits the Hall resistivity $\rho_{xy}$ and the magnetoresistance (MR) measured at a set of temperatures across the Curie point.  At $T$ = 160 K above $T_C$, neither $\rho_{xy}$ or MR show hysteresis. In contrast, hysteretic MR and remarkable THE signals are clearly captured below $T_C$, manifested as the hump-like features shaded in gray on each $\rho_{xy}$ curve. 
Before regarding THE as an indication of chiral spin textures, we rule out the possibility that the observed hump-like feature is originated from the summation of two opposite AHE signals for the following important factors: (1) The THE caused by two AHEs are often present at rather low temperatures and the overall line-shapes are highly sensitive to magnetic field,\upcite{Kan_2018_PRB,Wysocki_2020_PRM} whereas the THE of our (111) SrRuO$_3$ thin films is robust. (2) If the two-AHE scenario should be present, it would produce steps on the MR curves,\upcite {Wang_2020_npj,Jiang_2020_NM} which are absent in our case. (3) Our films are about 20 nm (corresponding to 87 u.c. along (111) orientation), such that the inhomogeneous issues induced by thickness or defects are practically negligible.\upcite{Kimbell_2022_CM} This is further corroborated by comparison to the behaviors of (111) SrRuO$_3$ films of the same thickness on SrTiO$_3$ substrates, where clean and conventional AHE signals are observed without any THE signature [Supporting Information Figure~S16]. 

To make a quantitative estimate, we extract the THE contribution $\rho_{\text{THE}}$ by taking the difference in $\rho_{xy}$ between upward and downward field scans.\upcite{Liu_PRL_2017, Jiang_2020_NM, Chen_2023_AFM} As shown in Figure~\ref{Fig2}b, the peak value of the THE humps ($\rho_{\text{THE, peak}}$) increases rapidly below the Curie temperature, reaching a maximum of $\sim$0.057 $\mu \Omega \cdot \text{cm}$ near 60 K, and tends to level off to $\sim$0.05 $\mu \Omega \cdot \text{cm}$ at the base temperature. Such a non-monotonic variation of $\rho_{\text{THE}}$ as a function of temperature is consistent with the THE behaviors reported in other systems such as CrTe/SrTiO$_3$\upcite{Zhao_2018_NR}, MnSi \upcite{Li_2013_PRL} and Tm$_3$Fe$_5$O$_{12}$/Pt.\upcite{Shao_2019_NE} Mapping of the $\rho_{\text{THE}}$ intensity versus field and temperature further demonstrates the THE signals are markedly observed below $T_C$ within a field range of 2.5 - 5.5 T [Figure~\ref{Fig2}c]. Considering the much narrower loop of magnetization with a coercivity of $\sim$0.6 T [Supporting Information Figure~S10-11], the presence of THE therefore plausibly signifies the formation of chiral spin states. 

\begin{figure}[htp]
\centering
\includegraphics[width=0.95\textwidth]{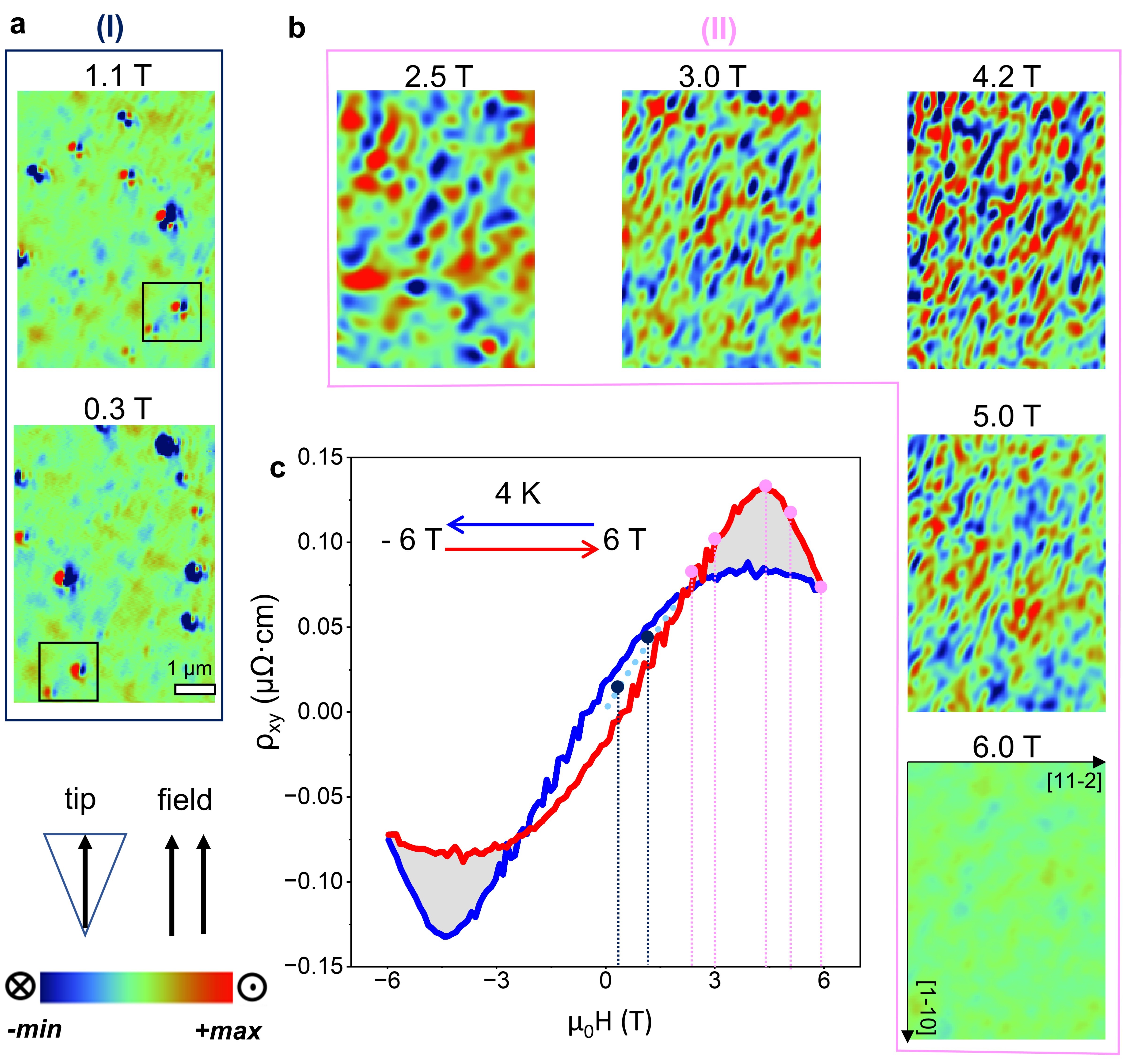}
\caption{\label{Fig3} 
{\bf Evolution of magnetic domains in (111) SrRuO$_3$ thin films at 4 K. (a)  MFM images recorded in regime (I) at 0.3 and 1.1 T on the initial magnetization process. The color bars stand for a frequency range from -1.0 Hz to 1.0 Hz, covering a total scanning area of 5$\times$7 $\mu$m. The outlined area on each figure in the black box highlights the splitting of the magnetic bubbles. (b)  MFM images recorded in regime (II) at a set of magnetic fields on the hysteretic branch with backward sweeping direction. The color bars stand for a frequency range from -0.35 Hz to 0.35 Hz. The crystallographic [1-10] and [11-2] directions are marked on the image of 6 T. (c) The full hysteresis loop of Hall resistivity at 4 K. The dark blue and pink circles refer to where the MFM scans were recorded in regime (I) and (II), respectively.}}
\end{figure}

Next, we perform MFM imaging at low temperatures to provide real-space insights \upcite{Li_2022_AM,Sohn_2021_AM,Rusu_2022_Nature} into the magnetic structures and shed light on the evolution of domains [Figure~\ref{Fig3}a-b]. In particular, in order to set up a comprehensive correspondence to the THE, the MFM images were recorded at a set of field spots in two regimes covering the entire $\rho_{xy}$ loop, as marked on Figure~\ref{Fig3}c. 
In regime (I) on the initial magnetization curve, there are magnetic bubble-like features observed at 0.3 and 1.1 T, which are randomly distributed on top of the overall weak contrast. Strikingly, the bubbles always appear in pairs with opposite out-of-plane magnetization components, which can be further split as field increases.      
In regime (II) on the THE hysteretic branch, however, stripe-like domains with interpenetrated opposite components appear at 2.5 T, which are more prominently seen at 3.0 and 4.2 T for sharper contrast, and become faded at 5.0 T. Eventually, at 6.0 T, the stripe domains diminish and the film is fully polarized into a single domain.    
It is noteworthy that the MFM experiments were conducted multiple times after thermal-field cycling and also at different temperatures [e.g, Supporting Information Figure~S5 for 10 K], giving rise to similar observations.  
These results therefore unravel the strip-like feature is reproducible and steady within its phase window, indicating a fundamental distinction of the magnetic states between the presence and absence of THE in our (111) SrRuO$_3$ films.    

\begin{figure}[htp]
\centering
\includegraphics[width=0.95\textwidth]{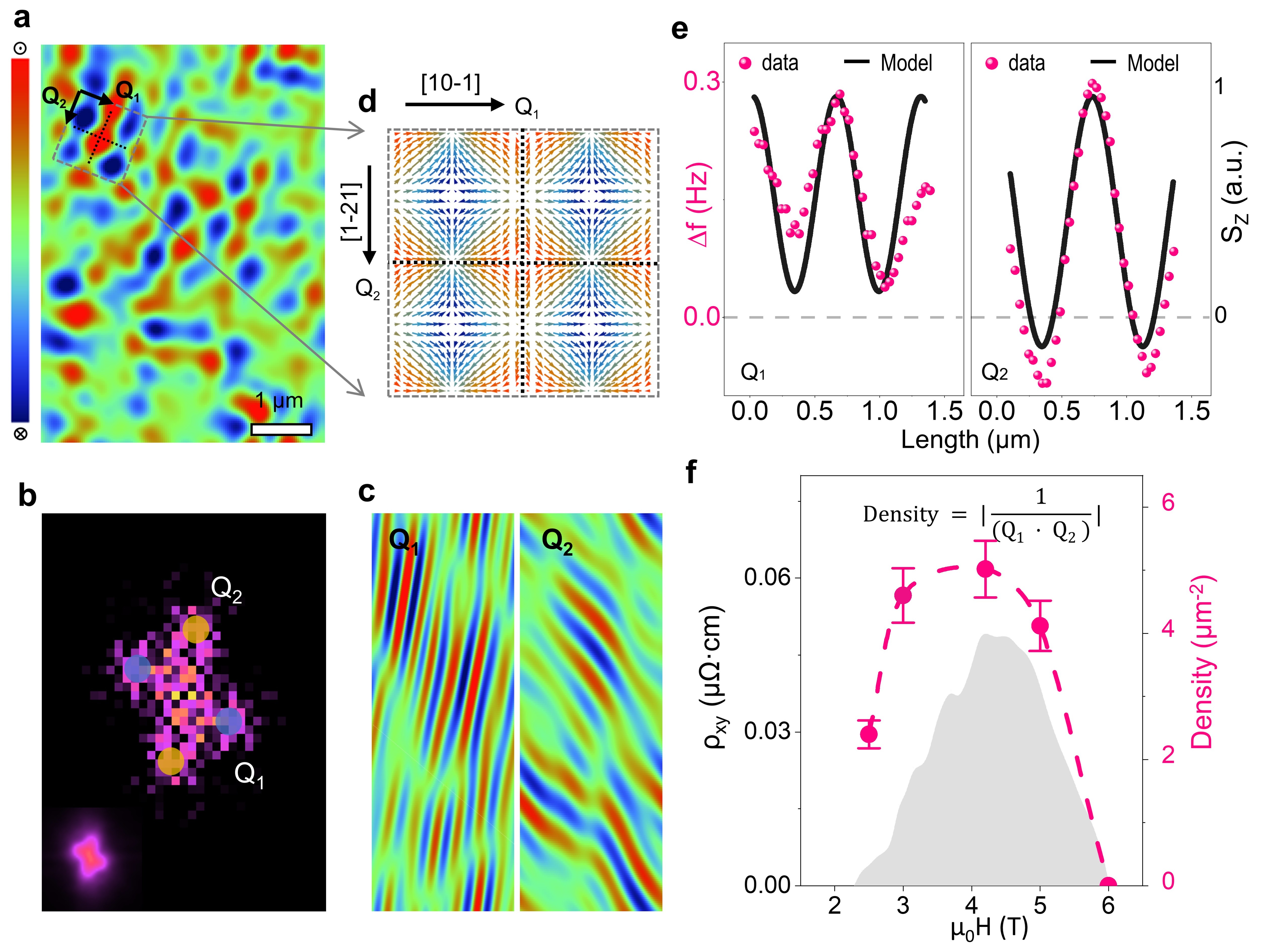}
\caption{\label{Fig4} 
{\bf Fourier analyses and model illustration of the double-$Q$ chiral spin crystal phase. (a) The 2.5 T MFM image after noise filter. (b)  Emergence of the double-$Q$ feature at 2.5 T on the FFT image, with the overall appearance shown at the left corner. The slight deviation from perfect orthogonality of $Q_1$ and $Q_2$ is due to the global averaging nature of FFT over the large scanning area. (c)  Isolated MFM images along each $Q$ direction by inverse FFT. (d) Simulated spin arrangements as a superposition of two orthogonal cycloidal spin spirals, projected along the out-of-plane direction. The propagation vectors $Q_1$ and $Q_2$ are approximately aligned along the crystallographic [10-1] and [1-21] directions. (e)  MFM data extracted from the line scans marked on (a)  and the corresponding spin z component from the line cuts on (d). (f) Plot of the Hall resistivity $\rho_{\text{xy}}$ and the density of the double-$Q$ spin crystal as a function of magnetic field at 4 K.}}
\end{figure}

More information about the domain characteristics is obtained from fast Fourier transform (FFT) analyses on the MFM images. While the FFT images in regime (I) are overall isotropic with one central peak, anisotropic patterns with two pairs of peaks emerge at 2.5 T, labeled as $Q_1$ and $Q_2$ as shown in Figure~\ref{Fig4}b, corresponding to real-space periodicities of $\sim$671$\pm$45 nm and $\sim$622$\pm$38 nm along the [10-1] and [1-21] directions, respectively. This leads to a periodic spin superstructure embedded on the lattice, referred to as a spin crystal phase.  
The isolated MFM images for each $Q$ vector obtained from inverse FFT clearly demonstrate the stripe-like domains are indeed manifested as the periodic modulations of the out-of-plane magnetization along two directions [Figure~\ref{Fig4}c]. Notably, the ``double-$Q$'' feature is concomitant to the THE, which persists to 5.0 T with varying magnitudes of the $Q$ vectors and finally vanishes at 6.0 T, when a saturated ferromagnetic domain is established [Supporting Information Figure~S7].

The noncoplanar spin configurations of the spin crystal phase can be reasonably simulated as a superposition of two orthogonal cycloidal spin spirals.\upcite{Chen_2016_SR} Specifically, the normalized local spin vector \(S(x, y)\) is expressed as: 
\begin{equation}
\begin{split}
& S(x, y) = I(x, y) \left[  \left( \sin(Q_yy), 0,  \cos(Q_yy) \right) +  \left( 0, \cos(Q_xx),  \sin(Q_xx) \right) \right] \\
& I(x, y) = \frac{1}{\sqrt{\left(  \sin(Q_yy) \right)^2 + \left(  \cos(Q_xx) \right)^2 +  \left(  \cos(Q_yy) + \sin(Q_xx \right)^2}}
\end{split}
\end{equation}
where \(I(x, y)\) is the normalization factor to ensure all spins of the same magnitude. $Q_x$ and $Q_y$ refer to the propagation vectors along $x$ and $y$ directions, whose values are determined by the FFT analyses. The resultant spin configurations projected along the $z$ direction are exhibited in Figure~\ref{Fig4}d. 
The variations of the MFM data extracted from the two orthogonal line scans drawn on Figure~\ref{Fig4}a are in accord with those from the simulated magnitudes of the spin z-component, as shown in Figure~\ref{Fig4}e.

It should be clarified that the spins do not form a perfect skyrmion crystal, as skyrmions require the full wrapping of magnetization onto a unit sphere, leading to the out-of-plane magnetization reversed completely at the center with respect to the boundary.\upcite{Tokura_2021_CR,Nagaosa_2013_NatNanotechnol,Kong_2018_APSCE}
Nevertheless, in analogy to the relationship between skyrmions and the THE, where the Hall resistivity is proportional to the number of skyrmions ($\rho_\text{THE} \propto n_\text{sk}$),\upcite {Wang_2022_PMS} the spin crystal phase also exhibits similar correspondence (i.e., $\rho_\text{THE} \propto D$, $D$ is the density of the noncoplanar spin textures, defined as $D = |1/(Q_1 \cdot Q_2)|$). This is visualized in Figure~\ref{Fig4}(f) by the simultaneously increasing and decreasing of $\rho_{xy}$ and $D$ as a function of magnetic field, suggesting the spin crystal phase with non-zero chiralities, giving rise to contributions to THE via the spin Berry phase. Notably, $Q_1$ remains nearly constant within experimental uncertainty, whereas $Q_2$ decreases with increasing magnetic field [Supporting Information Table~S1]. This distinct behavior likely arises from their differing magnetic anisotropy directions relative to the propagation vectors under a [111]-oriented field. For $Q_1$ (along [10-1]), the easy-axis anisotropy pins spin orientations, generating a strong restoring force that stabilizes its wavelength against field variations. In contrast,  $Q_2$ (along [1-21]) lies perpendicular to the easy axis, where weaker anisotropy competes with Zeeman energy, resulting in a field-dependent wavelength reduction.\\

Now we discuss the possible mechanism for the emergent chiral spin crystal phase in our (111) SrRuO$_3$ thin films. The formation of noncoplanar spin textures usually requires the presence of strong DM interaction, or dipolar interactions, and/or frustration induced by multiple magnetic interactions.\upcite{Rossler_2006_Nature,Leonov_2015_NC,Heinze_2011_NP,Ezawa_2010_PRL,Rohart_2013_PRB} Seddon {\it et al.} lately reported a similar incommensurate spin crystal phase in (001) SrRuO$_3$/PbTiO$_3$ bilayers, where the interfacial DM interaction significantly enhanced by the ferroelectric PbTiO$_3$ layer plays the decisive role in stabilizing the noncoplanar spin textures in 6 u.c. ultrathin SrRuO$_3$ layer.\upcite{Seddon_2021_NC} However, the interfacial DM interaction is unlikely the dominant factor in our case, due to the absence of any ferroelectric layer plus the much larger thickness of the films ($\sim$20 nm).\upcite{Zhang_2020_PRR} Alternatively, stabilization of the chiral spin crystal phase plausibly can rely on the presence of dipolar interactions and the delicate balance among multiple magnetic interactions.\upcite{Zhang_2020_PRR,Kwon_2012_JMMM,Scaramucci_2018_PRX,Ishiwata_2020_PRB,Mostovoy_2005_PRL}

To further examine this mechanism and pinpoint the critical parameters, Monte Carlo micromagnetic simulations\upcite{Sunny} were conducted on a buckled honeycomb lattice using the following Hamiltonian:\upcite{Lu_2020_NC}        
\begin{equation}
\begin{split}
\mathbf{H} = & \sum_{\langle i < j \rangle}(J_1 \, \mathbf{S}_i \cdot \mathbf{S}_j + \lambda_1 S_i^z S_j^z) 
+ \sum_{\langle \langle i < j  \rangle \rangle}(J_2 \, \mathbf{S}_i \cdot \mathbf{S}_j + \lambda_2 S_i^z S_j^z) 
+ \sum_{\langle \langle \langle i < j \rangle \rangle \rangle}(J_3 \, \mathbf{S}_i \cdot \mathbf{S}_j + \lambda_3 S_i^z S_j^z) \\ 
& + \sum_{i} A (S_i^z)^2 
+ E^\text{D-D} 
+ \sum_{i}\mathbf{B} \cdot \mathbf{S_i}
\end{split}
\end{equation}
Here the parameters $J_n$ and $\lambda_n$ ($n$ = 1, 2, 3) denote the magnitudes of the $n$-order nearest-neighbor isotropic and anisotropic exchange interactions, respectively; $A$ is the strength of the single-ion anisotropy; $E^\text{D-D}$ represents the dipole-dipole interactions between Ru ions. The effect of strain is equivalently included in these parameters. The calculations were run with sufficiently long steps until the complete convergence of energy.

\begin{figure}[htp]
\centering
\includegraphics[width=0.95\textwidth]{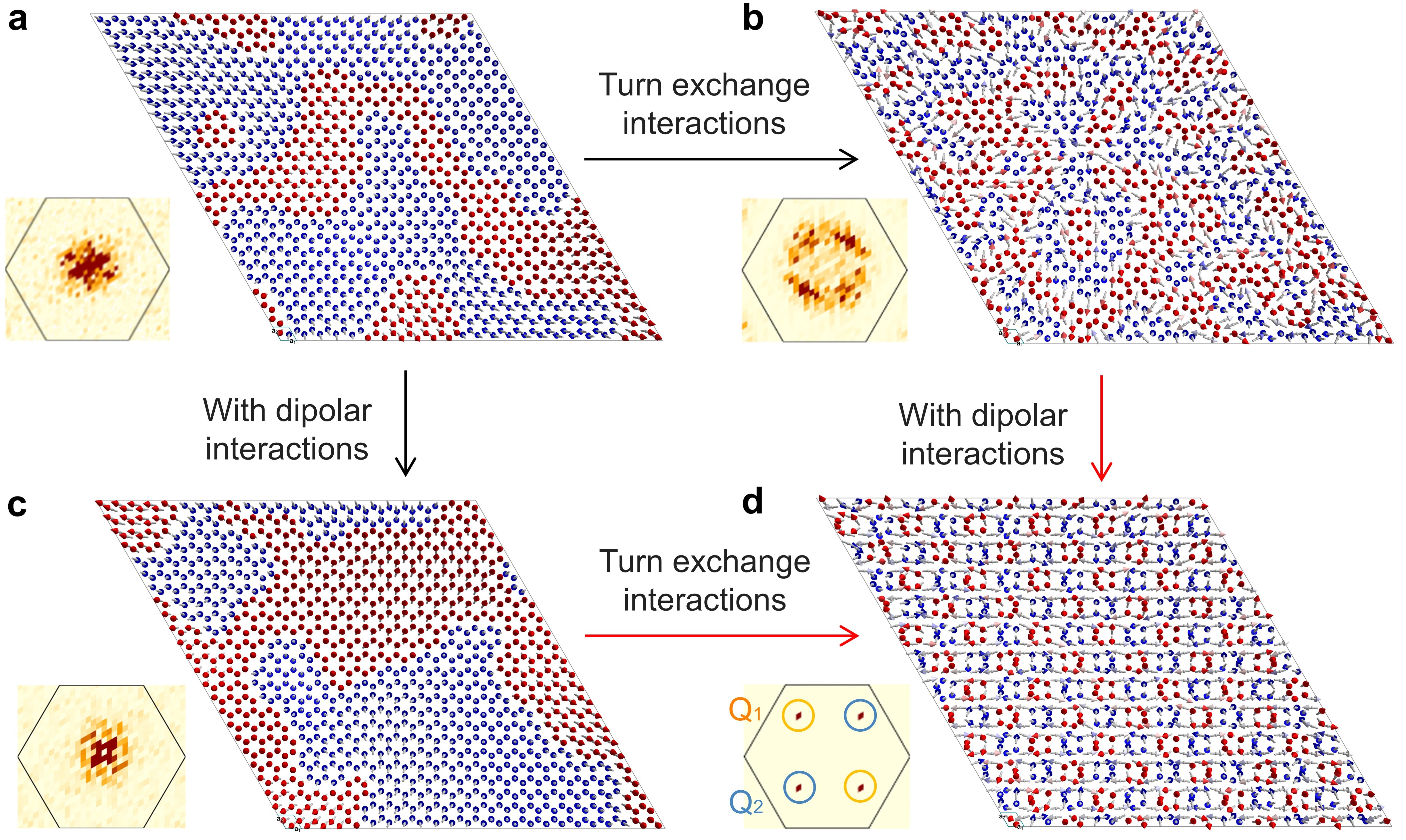}
\caption{\label{Fig5} 
{\bf Monte Carlo micromagnetic simulations on the buckled honeycomb lattice. The simulated spin structures are obtained in the cases of  (a) no dipolar interactions and bulk-like parameters;  (b) no dipolar interactions and manipulated exchange interactions;  (c) with dipolar interactions and bulk-like parameters;  (d)  with dipolar interactions and manipulated exchange interactions. The corresponding FFT images in each case are displayed at the bottom.}}
\end{figure}

It turns out that the formation of a periodic double-$Q$ noncoplanar spin state necessitates both the presence of dipolar interactions and manipulated exchange interactions.\upcite{Cenker_2022_NN,Yokota_2019_JPSJ}
Figure~\ref{Fig5}a shows the simulated spin structures in the absence of dipolar interactions, using the values of magnetic interactions (i.e., $J_n$, $\lambda_n$, $A$) from bulk SrRuO$_3$. This gives rise to typical ferromagnetic domain features and a central peak in its FFT image, testifying the validity of the model. Then we start exploring pathways to reach the spin crystal phase. On one hand, by tuning the exchange interactions, Figure~\ref{Fig5}b reveals the breakdown of large-scale ferromagnetic domains, replaced by short-range non-coplanar spin structures. However, no long-range superstructures with well-defined periodicities can be achieved at this stage. When the dipolar interactions are taken into account, the magnetic structure starts to exhibit periodic variations, eventually forming a periodic double-$Q$ noncoplanar spin state, as shown in the FFT image of Figure~\ref{Fig5}d. On the other hand, if maintaining the bulk parameters and only adding the dipolar interactions, it is yet not possible to reach the spin crystal phase, as illustrated in Figure~\ref{Fig5}c, where the overall ferromagnetic domain features do not change much. In particular, establishment of the final double-$Q$ feature requires a flip of sign of the second-order exchange anisotropy $\lambda_2$ [Supporting Information Table~4]. 


These findings highlight the physical origin of the ``double-Q'' chiral spin crystal phase. The parent ferromagnetic state remains robust, necessitating enhanced frustration to access a new energy minimum. This frustration originates from the cooperative interplay of dipolar interactions and the strain-induced reversal of $\lambda_{2}$. This specific parameter modification finds a physical basis in the strain-engineered modulation of Ru-O-Ru bonds, consistent with density functional theory (DFT) predictions of strain-driven easy-axis reorientation~\cite{Di_2021_PNAS}. 
Although our simulations utilized a 2D lattice, they capture the essential physics of the $\sim$20 nm films, where coherent tensile strain and strong interlayer coupling are expected to synchronize the magnetic texture across the thickness. Furthermore, while a weak strain-induced DMI may theoretically exist, it would serve primarily as a global chirality selector---lifting the degeneracy between enantiomeric states---rather than as the primary driving force, which we identify as the competition between dipolar interactions and geometric frustration.

\subsection{Conclusion}

In summary, we have achieved a robust chiral spin crystal phase in SrRuO$_3$ (111) epitaxial films, characterized by periodic noncoplanar spin arrangements along two orthogonal directions and robust THE. Micromagnetic simulations suggest its origin from the interplay of dipolar interactions and magnetic frustration, without the need of constructing substantial DM interactions via hetero-interface. Our findings thus present a novel route to realizing and manipulating the topological magnetic states in SrRuO$_3$ and offer a promising platform for designing perovskite-based spintronic devices.\\

\section{Experimental}
\subsection{Sample fabrication}
The (111) oriented SrRuO$_3$ thin films were grown on 5$\times$5 mm$^2$ (111)$_{pc}$ TbScO$_3$ single crystalline substrates by pulsed laser deposition. SrRuO$_3$ ceramic target was ablated using a KrF excimer laser ($\lambda$ = 248 nm, energy density $\sim$2 J/cm$^{2}$) with a repetition rate of 2 Hz. The deposition was carried out at a substrate temperature of 800 $^{\circ}$C, under an oxygen atmosphere of 75 mTorr. The films were post-annealed at the growth condition for 15 min, and then cooled down to room temperature. High-quality crystalline and single phase (111) SrRuO$_3$ thin film was eventually achieved. The whole deposition process was {\it{in-situ}} monitored by reflective-high-energy-electron diffraction. The synthesis of the (111) SrRuO$_3$ thin film on SrTiO$_3$ differs slightly from that on TbScO$_3$, requiring a growth temperature of 850$ ^\circ $C while maintaining an oxygen pressure of 75 mTorr.

\subsection{Magnetic force microscopy}
MFM measurements were performed with an Attocube-LT-SPM system with Liquid 2000 in the frequency modulation mode (FM-MFM) at T = 10 K and 4 K, respectively. The samples were mounted on a sample holder within a  chamber maintained at a high purity helium gas with the pressure of approximately 4000 Pa. Ultra-sharp tips (Nanosensor SSS-MFMR) with the tip remnant magnetization of 80 emu/cm$^{3}$ and elastic constant of 3.0 N/m was used at a resonant frequency of 70 kHz. A relatively high tip coercivity of 125 Oe. Before the MFM measurements, the probe was saturated using a permanent magnet for reliable and consistent MFM scans, enabling accurate investigations of the magnetic properties of the samples in our study. During the scans in the FM-MFM mode, the tip was lifted at a constant height of 60 nm above the sample surface, and the resonant frequency shift $\Delta f$ proportional to the magnetic force has been recorded as a function of the position. At a fixed temperature, following a 6 T field-cooling process with the magnetic field applied perpendicular to the film plane, MFM images were systematically acquired at selected field points along the ascending branch of the hysteresis loop. To ensure a comprehensive representation of the magnetization history, additional MFM imaging was conducted after zero-field cooling, with data collected at two distinct field points along the initial magnetization curve.

\subsection{Transport measurements}
The electrical transport measurements were performed in a Physical Property Measurement System (PPMS, Quantum Design) using the 4-point contact method. During the magneto-transport experiments, the magnetic field was applied along the [111] direction, and the current was driven along the [1$\bar{1}$0] direction. At each temperature, the longitudinal and transverse resistance were recorded while sweeping the field in a cycle from +6 T to -6 T to +6 T. Standard symmetrization (anti-symmetrization) treatment was further applied using these two branches of data on the longitudinal (transverse) resistance to achieve pure signals of the magnetoresistance (Hall resistance).

\subsection{Monte carlo micromagnetic simulations}
The Monte Carlo micromagnetic simulations were conducted using Sunny,\upcite{Sunny} an open-source Julia library that implements SU(N) spin dynamics for modeling atomic-scale magnetism. Details about the models, parameters, and fitting process are shown in the Supplementary Section II.

\begin{acknowledgement}

The authors deeply acknowledge A. Wu, H. Zhang, and L. Si for numerous insightful discussions, and the staff from BL07U beamline of Shanghai Synchrotron Radiation Facility (SSRF) for assistance on x-ray absorption spectroscopy data collection.This work is supported by the National Key R\&D Program of China (No. 2022YFA1403000, No. 2022YFA1403902), and the National Natural Science Foundation of China (No. 12204521, 12250710675, 12374155). A portion of this work was carried out at the Synergetic Extreme Condition User Facility (SECUF). A portion of this work was based on the data obtained at beamline 1W1A of Beijing Synchrotron Radiation Facility (BSRF-1W1A).\\ 

\end{acknowledgement}

\begin{suppinfo}

The following files are available free of charge.
\begin{itemize}
  \item Structural, transport and magnetic properties; Monte Carlo micromagnetic simulations; Auxiliary experiments: SrRuO$_3$ on SrTiO$_3$ (111).
\end{itemize}

\end{suppinfo}


\end{document}